# Fast switching dual Fabry-Perot cavity optical refractometry – methodologies for accurate assessment of gas density

Isak Silander,[1] Thomas Hausmaninger,[1] Michael Bradley,[1] Martin Zelan[2,a)], and Ove Axner[1,a)]

[1]*Department of Physics, Umeå University, SE-901 87 Umeå, Sweden*
[2]*Measurement Science and Technology, RISE Research Institutes of Sweden, SE-501 15 Borås, Sweden*

Dual Fabry-Perot cavity based optical refractometry (DFPC-OR) has previously demonstrated an extraordinary high potential for assessments of gas density. However, in practice, drifts of the FP cavities often limit its performance. By the use of two narrow-linewidth fiber lasers locked to two high finesse cavities, this work first shows, by the use of Allan-Werle plots, that DFPC-OR systems are predominantly affected by thermal drifts for "long" averaging times, while for "short" times, other types of noise dominates. Based on this, a novel strategy, termed fast switching DFPC-OR (FS-DFPC-OR), is presented in which measurements are made under such short times that the drifts of the cavity can be disregarded. A set of novel methodologies for assessment of both gas density and flow rates (in particular those from small leaks) that are not restricted by the conventional limitations imposed by the drifts of the cavity are presented. The methodologies that address gas density deal with assessments in both open systems and closed (finite-sized) compartments. They circumvent the problem with volumetric expansion, i.e. that the gas density in a measurement cavity is not the same as that in the closed external compartment that should be assessed, by performing a pair of measurements in rapid succession; the first one serves the purpose of assessing the density of the gas that has been transferred into the measurement cavity by the gas equilibration process, while the second is used to automatically calibrate the system with respect to the relative volumes of the measurement cavity and the external compartment. The methodologies for assessments of leak rates comprise triple cavity evacuation assessments, comprising two sets of measurements performed in rapid succession, supplemented by a third measurement at a certain time thereafter. This work provides also a clear explanation of why the technique has such an exceptionally small temperature dependence. It is concluded that FS-DFPC-OR constitutes a novel strategy that can be used for precise and accurate assessment of gas number density and gas flows under a variety of conditions, in particular under non-temperature stabilized ones. [Doc. ID XXXXX]

## I. INTRODUCTION

The most common means to detect the presence of gas, and quantitatively assess its amount, is to measure its pressure. However, although being a widespread methodology, it has some drawbacks. The most significant is that pressure is not solely proportional to the number of molecules in the gas, or its density, it is also a function of temperature.[1] This implies that it is not trivial to accurately assess its amount or density from a measurement of the pressure; any slight change (or drift) in temperature will affect the assessment noticeably. Other drawbacks are that pressure is not linearly related to gas density (due to a finite compressibility)[1] and that pressure gauges drift over time.

A means to alleviate this is to directly monitor the density of the gas.[2, 3] It has previously been shown that this can be done by measuring the refractivity of the gas.[2-14] Examples of advantages with this are that the density of a gas in a finite volume is independent of temperature, that refractivity is basically linear with gas density (it has solely a very weak non-linear dependence), and that this relationship does not have any temperature dependence.[1, 13-17]

Assessment of refractivity is normally done by interferometry and the highest precision is obtained by the use of optical refractometry (OR) based on Fabry-Perot (FP) cavities.[2, 3, 5-14, 17-22] The refractivity of the gas in the cavity is assessed by first locking the frequency of a laser to a longitudinal mode of a stable FP cavity and then monitoring the frequency of the laser light. Any change in the density of the gas, which gives rise to an associated alteration of its refractivity, can then be assessed as a change in the frequency of the laser beam.

However, although the leading order dependence of refractivity on gas density does not have any temperature dependence, the assessment of refractivity cannot be performed without a temperature influence. This is mainly due to thermal deformation of parts of the interferometer; in the case of a FP cavity, primarily the spacer between the mirrors. Although these types of effects can be held low by an intelligent choice of materials and the use of material treatment processes, they are in practice often the limiting factors in FP-Cavity based Optical Refractometry (FPC-OR).[11-13, 20, 22]

One means to minimize these types of effects is to use two FP cavities, referred to as a Dual Fabry-Perot Cavity based (DFPC) OR, here denoted DFPC-OR.[11, 13, 14, 19-23] In this case, one cavity is used as the measurement cavity while the other one is the reference cavity, thus eliminating the common effects of thermal drifts and relaxations of the spacer on the assessment. Besides minimizing effects of common-mode drifts, the introduction of a reference cavity also introduces a necessary reference to which the measured change in frequency of the measurement cavity can be directly assessed.

Unfortunately, it has been found, e.g. by Egan et al.,[13] that, despite the use of DFPC-OR, the measurements will, in general, unless delicate and meticulous curing and baking procedures have been used,[14] still be significantly affected by drifts, primarily those that originate from relaxations of the cavity spacer material and differential drifts of the lengths of the two cavities. This implies that it has been difficult to extract the full power of the DFPC-OR technique. There is therefore much to be gained if it would be possible to perform FPC-OR under drift-free conditions, i.e. so as to produce drift-free FPC-OR (here denoted DF-FPC-OR or DF-DFPC-OR). If this could be done, the technique would no longer suffer from any degradation of its precision and deterioration of its accuracy due to drifts and material relaxations. This leads to the question this work addresses: Given these practical conditions, how can DFPC-OR best be performed to allow for accurate assessment of gas number densities?

---

a) Author to whom correspondence should be addresses; electronic mail: martin.zelan@ri.se and ove.axner@umu.se.

It has earlier been shown by Silander et al., by the use of Allan-Werle plots, that the thermal drifts of a single FP cavity affect the assessment of refractivity by FPC-OR predominantly for "long" averaging times.[12] For "short" times, other types of noise, e.g. white or flicker noise, dominates.

It is shown in this work, by the use of a pair of narrow-linewidth fiber lasers that are locked to high finesse cavities, that the same holds for a DFPC-OR system. It is demonstrated that this system experiences a decreased Allan deviation of the detected beat signal from an empty DFPC-OR system for integration times up to around 0.5 s, after which it starts to increase, primarily due to drifts in the system. Based on this, it is suggested that the influence of drifts can be reduced, or even be made insignificant with respect to other sources of uncertainties in DFPC-OR, by performing measurements on short time scales; in practice, by performing measurements of the beat frequency of the two laser fields solely while the cavity is rapidly evacuated, henceforth referred to as *fast switching* DFPC-OR (FS-DFPC-OR). If such measurements (hereafter referred to as evacuation measurements) are performed solely within sufficiently short periods of time, gas densities can be assessed under drift-free conditions, and thus with no degraded uncertainty.

This methodology can directly be implemented to the cases when gas densities in an open system (i.e. a system of "infinite" size) are to be assessed. Based upon the general and explicit expressions for how a measured change in laser frequency is related to the density of a gas, which is derived in one of our accompanying works,[24] this work shows that the method can directly be used to assess gas density from an assessment of a change in beat frequency between two laser fields, each locked to a mode of its own cavity.

When densities in closed (i.e. finite-sized) compartments are to be assessed though, the assessment will be affected by volumetric expansion, i.e. the fact that a filling of the measurement cavity extracts a given fraction of the gas in the external compartment, given by the relative volumes of the measurement cavity and the external compartment. This means that the gas density in the measurement cavity will not be the same as the density the external compartment that should be assessed.

A methodology to circumvent this inconvenience is presented. It consists of performing two evacuation measurements in rapid succession. The first serves the purpose of assessing the density of the gas that has been transferred into the measurement cavity by the gas equilibration process, while the second is used to automatically calibrate the system with respect to the relative volumes of the measurement cavity and the external compartment, where the latter is done by measuring the decrease in density that comes with letting the gas in the closed external compartment diffuse into the measurement cavity a second time. By this, it is prophesized that gas densities in closed, finite-sized compartments can be assessed by FPC-OR with no need for any specific assessment of individual gas volumes or gas expansion factors, hence by means of volume-independent assessments.

In line with Hedlund and Pendrill,[9, 10] it is then proposed that also flows of gas can be assessed accurately by use of FS-DFPC-OR. In this case, a methodology is suggested that consists of three successive measurements. This methodology is outlined for two different situations: *i)* when the gas flow originates from a standard reference gas leak (assumed to have a constant leak rate) and *ii)* when it is caused by an orfice plate (for which the leak rate is assumed to decrease as the pressure in the compartment equilibrates with the surrounding).

Since the FS-methodologies presented here allow for assessments under drift-free conditions, they are useful for highly precise and accurate assessment of gas number density, and changes in such, under a variety of conditions; in particular the commonly occurring and important non-temperature stabilized ones.

Based on an accompanying work that derives general and explicit expressions for how a measured change in laser frequency in DF-DFPC-OR is related to the density of a gas,[24] the main aim of this paper is thus to present a basic experimental system and novel methodologies that can be used for assessment of gas density and changes of such under drift-free condition based on fast switching of gases, i.e. by use of FS-DFPC-OR, with significantly improved performance. The residual precision, accuracy, and temperature dependence of DF-DFPC-OR in general, and the FS-DFPC-OR methodologies that are portrayed in this work in particular, are estimated and discussed in some detail in another accompanying work.[25]

From all this, it is prophesized that by implementation of the fast switching methodology, the DFPC-OR technique can be utilized closer to its full power. It is therefore hoped that this work can serve as a basis for future designs and constructions of instrumentation that can benefit the most from the extraordinary power of DFPC-OR for gas density assessments.

## II. THEORY

### A. Expressions for gas number density in terms of the frequency shift of a laser locked to a mode of an optical cavity as the cavity is evacuated

*1. For the case with no relocking*

As is shown in our accompanying paper,[24] for DF-FPC-OR the density of gas in the cavity prior to a rapid evacuation, $\rho_{n,i}$, (where the subscript *i* stands for initial) can be related to the frequency shift of laser light, locked to a mode of the cavity, that follows an evacuation of the cavity, $\Delta \nu_l$, as

$$\rho_{n,i} = \rho_{n,f} + \frac{2}{3A_R} \tilde{\chi}_{\Delta\nu} \frac{\Delta \nu_l}{\nu_0} \left[ 1 + \tilde{B}_{\Delta\nu} \frac{\Delta \nu_l}{\nu_0} + \tilde{C}_{\Delta\nu} \left( \frac{\Delta \nu_l}{\nu_0} \right)^2 + \ldots \right], \quad (1)$$

where $\rho_{n,f}$ represents the residual density of the gas after the evacuation (where *f* stands for final, to accommodate for the case that not all gas is removed from the cavity by the evacuation process), henceforth referred to as the residual gas density, $A_R$ is the (number or molar) polarizability of the gas, $\tilde{\chi}_{\Delta\nu}$ is a dimensionless factor that is close to unity but includes contributions from a number of entities (e.g. the instantaneous deformation of the spacer, $\varepsilon$, the dispersion of the mirrors and the gas, $\varsigma$ and $\eta$, and the refractivity of the residual density of the gas, $n_f - 1$, henceforth referred to as the residual refractivity), $\nu_0$ is the frequency of the laser light when locked to the cavity mode addressed in an evacuated cavity, while $\tilde{B}_{\Delta\nu}$ and $\tilde{C}_{\Delta\nu}$ are normalized coefficients that, together with $\tilde{\chi}_{\Delta\nu}$, are given in Table 2 in Ref. [24], expressed in terms of the $A_R$, $B_R$, and $C_R$ virial coefficients as well as the $\varepsilon$, $\varsigma$, and $\eta$ parameters. All entities are properly defined in Ref. [24]. For the case with no relocking, $\Delta \nu_l$ can be interpreted as the shift of the beat frequency between the two laser fields that follows an evacuation of the cavity, $\Delta \nu_{beat}$.

*2.A.2. For the case with relocking*

For the case when the laser is relocked to another cavity mode during the evacuation, differing in mode number by $\Delta q$, $\rho_{n,i}$ is given by

$$\rho_{n,i} = \rho_{n,\mathrm{f}} + \frac{2}{3A_R}\tilde{\chi}_{\Delta q}\left\{\frac{\Delta q}{q_0}\left[1 + \tilde{B}_-^{(\Delta q)^2}\frac{\Delta q}{q_0} + \tilde{C}_-^{(\Delta q)^3}\left(\frac{\Delta q}{q_0}\right)^2\right]\right.$$

$$+ \frac{\Delta \nu_l}{\nu_0}\left[\tilde{A}_{\Delta\nu}^- + \tilde{B}_{\Delta\nu}^{\Delta q}\frac{\Delta q}{q_0} + \tilde{C}_{\Delta\nu}^{(\Delta q)^2}\left(\frac{\Delta q}{q_0}\right)^2\right] \quad (2)$$

$$\left. + \left(\frac{\Delta \nu_l}{\nu_0}\right)^2\left[\tilde{B}_{(\Delta\nu)^2}^- + \tilde{C}_{(\Delta\nu)^2}^{\Delta q}\frac{\Delta q}{q_0}\right]\right\}$$

where $q_0$ is the mode number of the cavity mode addressed originally by the laser field and the $\tilde{\chi}_{\Delta q}$, $\tilde{B}_-^{(\Delta q)^2}$, $\tilde{C}_-^{(\Delta q)^3}$, $\tilde{A}_{\Delta\nu}^-$, $\tilde{B}_{\Delta\nu}^{\Delta q}$, $\tilde{C}_{\Delta\nu}^{(\Delta q)^2}$, $\tilde{B}_{(\Delta\nu)^2}^-$, $\tilde{C}_{(\Delta\nu)^2}^{\Delta q}$, and $\tilde{C}_{(\Delta\nu)^3}^-$ coefficients, which again incorporate the virial coefficients, the $\varepsilon$, $\varsigma$, and $\eta$ parameters, and the residual refractivity, are explicitly defined in Table 3 in Ref. [24].

The Eqs. (1) and (2) are thus the basis for assessments of gas density from a measurement of the shift of the cavity mode addressed by the laser light that follows an evacuation of the cavity in DF-FPC-OR systems and therefore also in the FS-DFPC-OR systems discussed in this work.

## III. EXPERIMENTAL SET-UP

To assess the typical characteristics of a DFPC-OR system (in particular its detection sensitivity for various signal averaging times), a prototype FPC-OR system, based on a dual-cavity design consisting of a measurement and a reference cavity bored in the same cavity spacer, was constructed. Two independent narrow linewidth fiber lasers were used, one locked to each cavity, and the difference in frequency of the two was measured directly as a beat note by a large bandwidth photodetector. Prior to the measurements, the cavity spacer had not been exposed to any thermalization or curing to decrease internal relaxations.

The experimental setup is illustrated in Fig 1. The light from two Er-doped fiber lasers (EDFL, Koheras Adjustik E15), with wavelengths of 1.53 μm, was sent through fiber-coupled acousto-optic modulators (AOM, AA Opto-Electronic, MT110-IR25-3FIO). The first order output of each AOM, shifted by 110 MHz, was sent through a 90/10 polarization maintaining fiber splitter (Thorlabs, PMC1550-50B-FC), whose high transmission output was connected to an electro-optic modulator (EOM, General Photonics, LPM- 001-15) modulated at 20 MHz for the purpose of Pound-Drever-Hall (PDH) locking to the cavities. Each beam was then coupled into free space by an output coupler and sent through mode-matching lenses, a polarizing beam splitter (PBS), and a quarter-wave (λ/4) plate, before it entered the cavity. Each cavity was made of high reflectivity mirrors, glued to a spacer made of Zerodur, yielding a finesse of ~10 000 and an FSR of ~800 MHz.

The two cavities were drilled in a common Zerodur block with a squared cross section (100 by 100 mm) and a length of 188 mm. To minimize the influence of instantaneous length deformations, the cavities were made with a diameter significantly smaller than that of the mirrors; 5 vs. 25 mm.

For the locking, the back-reflected light from each cavity, which again passed through the quarter-wave (λ/4) plate, was reflected by the PBS onto a large bandwidth photo-detector (New Focus, 1611). The output of each detector was demodulated at 20 MHz to produce an PDH error signals. The slow components (<100 Hz) of each PDH error signal were sent through custom made electronics to the piezoelectric transducer in the corresponding fiber laser, while the fast components (>100 Hz) were fed through a commercial PID controller (Toptica, FALC 110 for one of the cavities and New Focus LB 1005 for the other) to a voltage controlled oscillator (VCO), connected to the corresponding AOM. In order to monitor the beat note of the two lasers, the low transmission outputs of the 90/10 fiber splitters were combined in a 50/50 fiber combiner and sent to a fiber-coupled large bandwidth photo-detector (New Focus, 1611). The frequency of the beat note was measured with a frequency counter (Agilent, 53230A) with a time resolution down to 1 μs.

To stabilize the temperature of the dual-cavity block, the latter was placed directly onto the optical table, which, in turn, was placed inside a temperature stabilized room. The temperature drift of the cavity spacer during measurements cycles was assessed to be, on average, around 0.02 °C/h.

The cavities were connected to a common gas supply and vacuum system (not shown in Fig. 1) that allowed for controlled changes of gas in the cavities. The pressure inside the cavities was monitored continuously with two pressure gauges (Oerlikon Leybold Vacuum, Ceravac CTR91-1Torr/1000Torr).

## IV. RESULTS

### A. Measurement of the frequency of the beat note

In order to characterize the system, the beat note between the two fiber lasers was measured while both cavities were actively evacuated. Figure 2 displays, some typical beat note data, i.e. $\nu_{beat}(t)$, defined as the difference between the frequencies of the two laser fields that are locked to the measurement and the reference cavities, respectively, taken under dead-time-free measurement conditions with four different gate times of the frequency counter, 1 μs, 10 μs, 100 μs, and 1 ms, respectively. Since both cavities were evacuated, none of them were considered containing any gas. The scattering and drifts of the data was therefore presumed to originate solely from the measurement system, predominantly the linewidths of the lasers under locked conditions, i.e. $\sigma_{\Delta\nu}$, and the drift of the difference in physical length of the two cavities. Contributions from the photodetector and the frequency counter were considered negligible.

### B. The Allan deviation of the frequency of the beat note

The beat note data were analyzed by the use of the Allan deviation. The Allan deviation constitutes the square root of the Allan variance, which, in turn, is defined as one half of the time average of the squares of the differences between successive readings of the entity sampled over a given sampling period.[26] Since the Allan deviation is a function of the averaging time, commonly denoted $\tau$, it is displayed as a graph (in log-log format) rather than a single number. It is a particularly useful tool for assessments of sources of noise in measurement system. A low Allan variance for a certain averaging time is a characteristic of a system with good stability over that particular measured period.

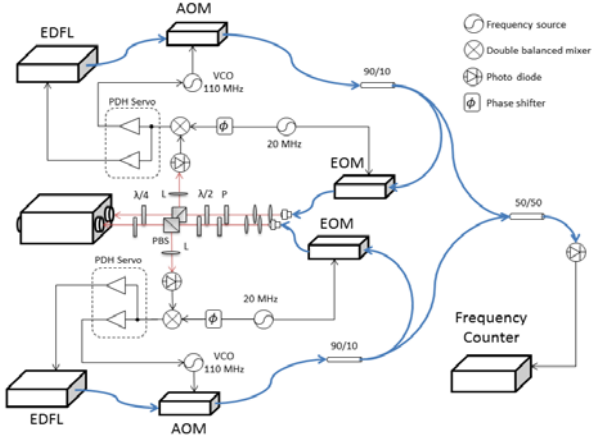

FIG 1. (Color online) Experimental setup of the prototype FPC-OR system. Two narrow linewidth fiber lasers (EDFL) are locked by the Pound-Drever-Hall technique to individual Fabry-Perot cavities mounted on the same Zerodur spacer. One of the cavities contains the gas whose density is to be assessed, while the other one constantly evacuated, acts as a reference. Assessments are done by measuring the change in the difference in frequency of the two lasers while the measurement cavity is evacuated by detecting their beat note with a photodetector and assessing its frequency by a frequency counter. See text for details. Blue curves with arrows indicate optical fibers and the direction of light propagation, red lines with arrows indicate free space propagation of light, and black lines and arrows indicate electric wires and signals.

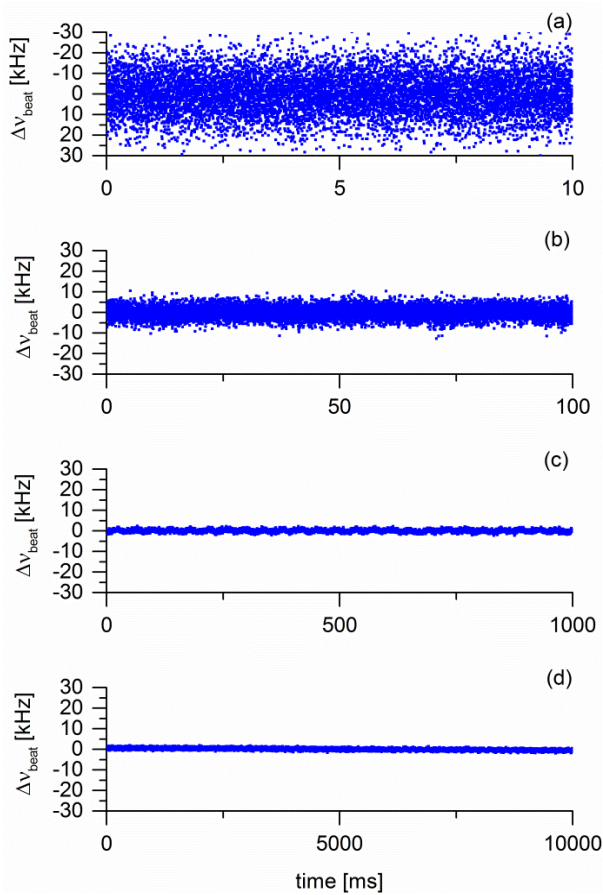

FIG 2. Frequency of the beat note between the two lasers as a function of time for four different frequency counter gate times, panel a – d: 1 µs, 10 µs, 100 µs, and 1 ms, respectively. For comparison, all data sets have been offset to an initial beat note frequency of zero. For the shortest gate times (the panels a and b), the noise is due to the finite linewidth of the lasers, while for the longest (the panels c and d), the data is significantly affected by cavity drifts.

The Allan deviation of the measurements displayed in Fig. 2 is shown in Fig. 3 with the four gating times marked with different types of solid curves: 1 µs, red curve; 10 µs, blue curve; 100 µs, green curve; and 1 ms, black curve. For comparison, the Allan deviation of the frequency of the beat note between a laser locked to a single FP cavity and a frequency comb (data from Ref. [12]) is shown by the black dashed curve.

Figure 3 shows first of all that although both setups are affected by drifts for long averaging times, the dual-cavity system is superior to the single FP cavity system for short averaging times (in this case up to 10 s). A closer analysis indicates that for the shortest averaging times (below 1 ms), for which the system has a white noise behavior that amounts to ~ 10 Hz s$^{1/2}$, the Allan deviation originates from the combined linewidths of the two lasers, $\sigma_{\Delta\nu}$. There is a plateau between 1 and 10 ms, which is assumed to be caused by the handover between the laser and the AOM servos in the locking of the lasers to their cavities. This indicates that the data in this interval are affected by technical noise from the instrumentation. For yet longer averaging times, up to around a fraction a second, the Allan deviation continues to decrease with averaging time.

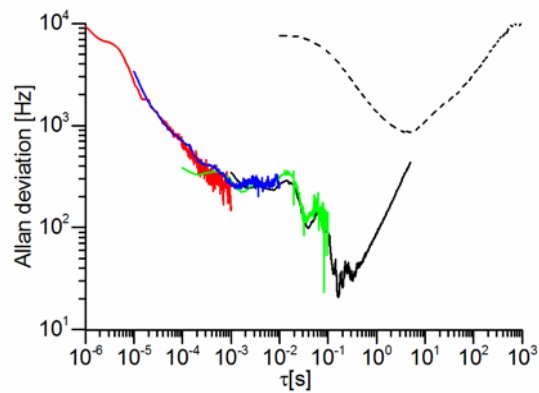

FIG. 3. The Allan deviation of the frequency of the beat note of the measurements displayed in Fig. 2, $\sigma_{\Delta\nu}$, as a function of averaging time, $\tau$, for four different gating times: 1 µs, red solid curve; 10 µs, blue solid curve; 100 µs, green solid curve; and 1 ms, black solid curve. The Allan deviation of the frequency of the beat note between a laser locked to a single FP cavity and a frequency comb from our first paper[12] is shown by the black dashed curve.

For longer averaging times (above 0.5 s), the sensitivity is limited by the stability of the frequencies of the cavity modes, $\sigma_{\nu_q}$, which drift at a combined rate of ~ 75 Hz s$^{-1}$. This implies that for such measurement times the relative stability of the cavity modes, defined as $\sigma_{\nu_q}/\nu_0$, can be estimated to be ~ 4 × 10$^{-13}$ $t$ s$^{-1}$. This is considered to originate from cavity mode creeping of the unthermalized piece of Zerodur from which the DFPC was made.

Even though some of the data are affected by technical noise, Fig. 3 shows that the frequency of a cavity mode can be assessed, for a range of measurement times, in this case, between 0.1 and 2 s, with a precision below 100 Hz, and that the system exhibits a minimum Allan deviation of the beat note of 36 Hz (for an integration time of 0.25 s).[27,28]

Although all this indicates both that the lasers can be locked to cavity modes with a precision of a fraction of the width of the cavity modes (to a relative precision of 2 × 10$^{-13}$) and that the performance of the system can be further improved by a more carefully constructed locking process, in particular for averaging times in the ms range, most importantly, it demonstrates that assessments that are not affected by cavity drifts or relaxations can be performed if the measurements are performed under sufficiently short measurement times. Hence, although the system used in this study shows that drifts set-in already from a measurement time of around 0.5 s, it is assumed that systems with less drifts, i.e. in which the drifts set in at longer times, can be constructed by a more careful choice of cavity spacer material and mirror substrate, mounting of mirrors, and material treatment procedures (e.g. curing and out-baking). One such example is the DFPC-OR system realized by Khelifa et al. which has demonstrated white-noise limited behavior to at least 100 s.[5, 20]

However, to reduce the influence of drifts, so as to eliminate the degradation of the precision and reduce the deterioration of the accuracy of the technique, and to in part remedy the need for advanced cavity spacer treating processes, based on our characterizing assessments described above, we present below methodologies for FS-DFPC-OS that can address assessments of gas density and changes in such under a variety of conditions, in particular for assessments of gas density in both open systems and closed compartments and for measurements of flows of gases from both standard reference gas leaks and orfice plate. The suggested methodologies assume access to instrumentation that can accommodate fast gas switching, based on measurements performed on cavities during evacuations in rapid succession, under such a short time that cavity drifts and relaxations will not significantly affect the measurements. In order to additionally improve on the accuracy of the technique, the methodologies presented are based on an extended theoretical description of DF-DFPC-OR that is given in one of our accompanying papers[24] and shortly summarized above, by the Eqs (1) and (2). By this, the methodologies are expected to serve as a basis for design of how such future high performance DF-FFPC-OR instrumentation could be constructed.

## V. SUGGESTED EXPERIMENTAL SYSTEM AND METHODOLOGIES FOR DRIFT-FREE DFPC-OR BY FAST SWITCHING DFPC-OR

Based on the favorable properties of the DF-DFPC-OR technique, it is here suggested how a FS-DFPC-OR system could be realized and how such a system could be used to provide assessments of gas density and leak rate with high precision and accuracy under some typical conditions.

### A. The generic FS-DFPC-OR system – Experimental systems and realizations

Consider in general a DFPC-OR system that consists of two FP cavities that can, but do not need to, be bored in the same block of cavity spacer, as is shown in Fig. 4. Consider also two tunable narrow line width lasers, indicated by the red arrows in the same figure, each locked to a longitudinal mode of one of the cavities by the PDH technique, as illustrated in Fig. 1. Let us more specifically envision a FS-DFPC-OR system in which gas is let into one of the cavities, cavity 1, from an open system or closed chamber whose gas density, $\rho_{ext}$, is to be assessed, while the other, cavity 2, is evacuated and serves as the reference cavity. Let us furthermore assume that there are at least two valves in the system, one between the external system/compartment and cavity 1, $X_{1,ext}$, and one between cavity 1 and the vacuum pump, $X_{1,vac}$. Although not entirely necessary, it is advisory to also have a valve between cavity 2 and the vacuum pump, in the figure denoted $X_{2,vac}$. Let us, for simplicity, denote this combination of spacer block, cavities, lasers, and valves the "generic FS-DFPC-OR system".

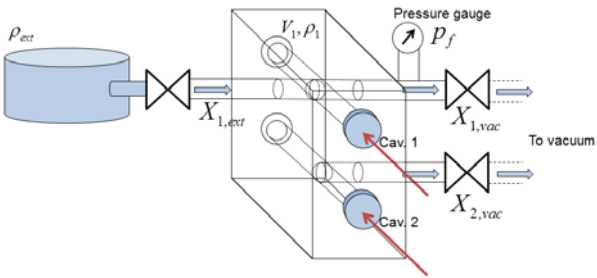

FIG 4. Schematic illustration of a possible realization of a FS-DFPC-OR system, here referred to as the generic FS-DFPC-OR system. Cavity 1 is connected to an open system or closed chamber whose gas density, $\rho_{ext}$, is to be assessed, while the other, cavity 2, is evacuated and serves as the reference cavity.

By monitoring the beat frequency of the two lasers, changes in the refractivity of the gas in cavity 1 can be assessed swiftly and with high precision with respect to that in cavity 2 under various conditions.

As is illustrated by the three panels of Fig. 5, this work will specifically focus on three types of applications: panel a): when gas density in an open system is to be assessed, let cavity 1, whose volume is $V_1$, be directly attached to either an open system; panel b): when the gas density from a closed external compartment is to be assessed, let the same cavity be connected to a closed external gas compartment, and panel c): when leak rates are to be assessed, let it be connected to a closed compartment to which gas leaks are coupled.

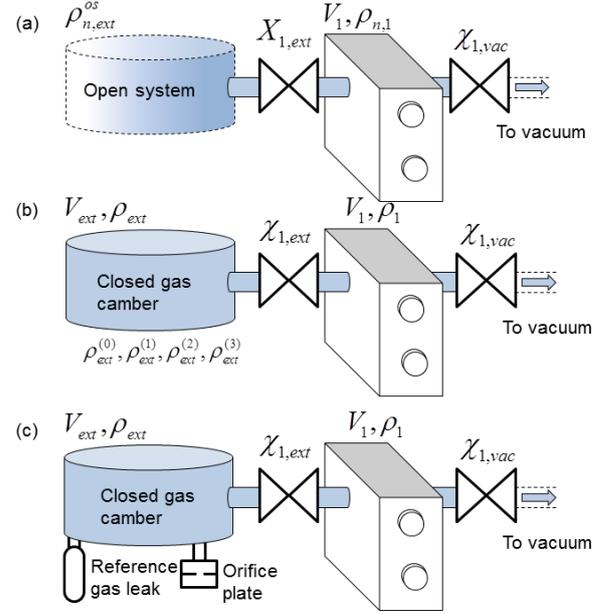

FIG 5. Three possible uses of the generic FS-DFPC-OR system. Panel a) Suggested use of the system for assessment of gas density in an open (i.e. infinitely large) external system. Panel b) Suggested use of the system for assessment of gas density in a finite-sized system, represented by an external compartment with a given volume $V_{ext}$. Panel c) Suggested use of the system for assessment of gas leaks, either a standard reference gas leak or an orfice plate. The gas density in the open system is denoted $\rho_{n,ext}^{os}$, that in the external compartment is termed $\rho_{n,ext}$, while that of cavity 1 is termed $\rho_{n,1}$. The original density in the external compartment, which is to be assessed, is denoted $\rho_{ext}^{(0)}$. After each filling of cavity 1 from a closed external compartment, the density in that compartment will change (be decreased) and be denoted by $\rho_{n,ext}^{(1)}$, $\rho_{n,ext}^{(2)}$, and $\rho_{n,ext}^{(3)}$, where the superscript $^{(i)}$ indicates how many evacuations of the external gas compartment (fillings of cavity 1) that has been performed. The corresponding number (or molar) densities in cavity 1 are denoted $\rho_{n,1}^{(1)}$, $\rho_{n,1}^{(2)}$, and $\rho_{n,1}^{(3)}$, respectively.

### B. Methodology for assessment of gas density in an open system

As is shown in Fig. 5a, when the density in an open system is to be assessed, the measurement cavity of the generic FS-DFPC-OR system is directly connected to the open system by a valve, $X_{1,ext}$. This valve is then opened long enough so that the conditions in the measurement cavity are allowed to equilibrate with those in the open system. In this case, since this system is infinitely large, the density of gas in the measurement cavity, $\rho_{n,1}^{(1)}$,[29] will be the same at that in the open system, $\rho_{n,ext}^{os}$, i.e.

$$\rho_{n,1}^{(1)} = \rho_{n,ext}^{os}. \tag{3}$$

When this has been achieved, the valve is closed.[30]

While the beat frequency between the two laser beams is monitored, the cavity is then rapidly evacuated by opening valve $X_{1,vac}$. As long as the reference cavity is held at vacuum, and the assessment is made fast enough so drifts can be neglected, the change in beat frequency between the two lasers, $\Delta v_{beat}$, corresponds to the change in the frequency of the mode of cavity 1 that is addressed, $\Delta v_l$, which, according to one of our accompanying papers,[24] in turn, is predominantly due to the change in gas density.

Based on the description of how the gas density depends on the measured beat frequency that was given in section 2, and the information given in Ref. [24], as is conveyed in Ref. [25], it is possible to conclude that the density of gas in the open system can, for a well characterized system, be assessed with excellent precision and good accuracy. For the case when no relocking is used, the Eqs. (1) and (3) then show that it is possible to assess $\rho_{n,ext}^{os}$ from the measurement of $\Delta v_l^{(1)}$ according to

$$\rho_{n,ext}^{os} = \rho_{n,1}^{(1)} = \rho_{n,1,i}^{(1)}$$
$$= \rho_{n,1,f}^{(1)} \qquad (4)$$
$$+ \frac{2}{3A_R} \tilde{\chi}_{\Delta\nu} \frac{\Delta\nu_l^{(1)}}{\nu_0} \left[ 1 + \tilde{B}_{\Delta\nu} \frac{\Delta\nu_l^{(1)}}{\nu_0} + \tilde{C}_{\Delta\nu} \left( \frac{\Delta\nu_l^{(1)}}{\nu_0} \right)^2 + \ldots \right],$$

where the notation $\rho_{n,1,i}^{(1)}$ and $\rho_{n,1,f}^{(1)}$ is used for $\rho_{n,1}^{(1)}$ before and after evacuation, respectively.

When relocking is used, $\rho_{n,ext}^{os}$ can be assessed by the expression

$$\rho_{n,ext}^{os} = \rho_{n,1}^{(1)} = \rho_{n,1,i}^{(1)}$$
$$= \rho_{n,1,f}^{(1)} + \frac{2}{3A_R} \tilde{\chi}_{\Delta q} \left\{ \frac{\Delta q^{(1)}}{q_0} \left[ 1 + \tilde{B}_{-}^{(\Delta q)^2} \frac{\Delta q^{(1)}}{q_0} + \tilde{C}_{-}^{(\Delta q)^3} \left( \frac{\Delta q^{(1)}}{q_0} \right)^2 \right] \right.$$
$$+ \frac{\Delta\nu_l^{(1)}}{\nu_0} \left[ \tilde{A}_{\Delta\nu}^{-} + \tilde{B}_{\Delta\nu}^{\Delta q} \frac{\Delta q^{(1)}}{q_0} + \tilde{C}_{\Delta\nu}^{(\Delta q)^2} \left( \frac{\Delta q^{(1)}}{q_0} \right)^2 \right] \qquad (5)$$
$$\left. + \left( \frac{\Delta\nu_l^{(1)}}{\nu_0} \right)^2 \left[ \tilde{B}_{(\Delta\nu)^2}^{-} + \tilde{C}_{(\Delta\nu)^2}^{\Delta q} \frac{\Delta q^{(1)}}{q_0} \right] \right\},$$

where $\Delta q^{(1)}$ is the number of modes the laser field jumps in the relocking process during the first evacuation of the measurement cavity.

Equation (4) shows that to assess the density of the gas in the open system, $\rho_{n,ext}^{os}$, from a measured shift in frequency, $\Delta\nu_l^{(1)}$, one needs to have access to a number of entities, primarily the (number or molar) polarizability of the gas addressed, $A_R$, but also the frequency of the laser when locked to the cavity mode in an evacuated cavity, $\nu_0$, the $\tilde{B}_{\Delta\nu}$ and $\tilde{C}_{\Delta\nu}$ coefficients, and the dimensionless factor, $\tilde{\chi}_{\Delta\nu}$ (all defined above). As is discussed in more detail in Ref. 25, all these entities should therefore be assessed either by a separate characterization procedure or be taken from the literature. When this has been done, the main contributions to the finite accuracy of the technique have been eliminated.

In addition, since an evacuation will not necessarily remove all gas from the cell, the assessment will also be affected by the residual gas density in the measurement cavity, $\rho_{n,1,f}^{(1)}$. However, since this depends on the procedure used, it cannot be assessed by a general calibration procedure beforehand. One the other hand, as is discussed in more detail in Ref. 25, it can, if needed, and as is indicated in Fig. 4, be assessed by a pressure assessment, e.g. by the use of a pressure gauge. Since this entity solely provide a small contribution to the final assessment, it is sufficient if $\rho_{n,1,f}^{(1)}$ is assessed from a measurement of the pressure, i.e. $p_{1,f}^{(1)}$, by the use of the ideal gas law, for the case when $\rho_{n,1,f}^{(1)}$ is expressed as a number density, by $\rho_{n,1,f}^{(1)} = p_{1,f}^{(1)}/kT$. It is shown in Fig. 2a in Ref. 25 that it suffices to reduce the pressure to 13 Pa (0.1 Torr) to be able to assess the residual gas density with a pressure gauge (CTR 100 0.1, Leybold) that provides a precision in the assessment of gas density of $10^{-6}$ mol/m$^3$, which, for assessment under STP conditions, corresponds to a relative precision of $2 \times 10^{-8}$.

For the case with relocking, the expressions above shows that the experimentalist also needs to know how many modes the laser has jumped, $\Delta q^{(1)}$. To obtain the highest accuracy, it would also be advantageous if $q_0$ could be assessed with no error. Both these are possible to achieve. As has been discussed by Egan et al.,[13] the latter can be assessed by knowledge about the FSR and the use of a wavelength meter with a resolution that is at least half of the FSR. The former can be assessed by monitoring the residual frequency shift of the beat note or by use of the wavelength meter.

The reader is referred to Ref. 25 for a further discussion about the precision, accuracy, and temperature dependence of this methodology.

## C. Methodology for assessment of gas density from a closed external compartment

As is shown in Fig. 5b, also for the case when the gas number density in a finite-sized closed external compartment, $\rho_{n,ext}^{(0)}$, is to be assessed, the generic FS-DFPC-OR system can be directly connected to this compartment. In this case though, a double assessment methodology should preferably be performed in which the first measurement serves the purpose of assessing the density of the gas that has been transferred into the measurement cavity by the gas equilibration process, while the second is used to automatically calibrate the system with respect to its ratio of volumes.

*1. Single refractivity assessment*

The gas density in the external compartment is thus first to be analyzed by a single measurement by the same type of cavity evacuation procedure as was used for the open system above. The measurement procedure therefore starts by evacuating cavity 1. The gas in the external compartment is then expanded into the measurement cavity by an opening of valve $X_{ext,1}$. When the gas conditions in the two cavities have equilibrated, the gas densities in the two compartments, $\rho_{n,ext}^{(1)}$ and $\rho_{n,1}^{(1)}$, are again the same, but are in this case related to the gas density to be assessed, $\rho_{n,ext}^{(0)}$, by

$$\rho_{n,1}^{(1)} = \rho_{n,ext}^{(1)} = \frac{V_{ext}}{V_{ext} + V_1} \rho_{n,ext}^{(0)} = \frac{1}{\chi_V} \rho_{n,ext}^{(0)}, \qquad (6)$$

where $(V_{ext} + V_1)/V_{ext}$, hereafter denoted $\chi_V$, will be referred to as the gas expansion factor. This implies that the gas density in the measurement cavity, $\rho_{n,1}^{(1)}$, is solely a fraction of the initial density of the external gas compartment, $\rho_{n,ext}^{(0)}$, given by the inverse of the gas expansion factor.[31]

Equation (6) then indicates, together with Eq. (1), that the gas density to be assessed, *viz.* that in the external gas compartment prior to any gas equilibration, $\rho_{n,ext}^{(0)}$, can, in the case with no relocking, be determined from the measured value of $\Delta\nu_1^{(1)}$ according to[32]

$$\rho_{n,ext}^{(0)} = \chi_V \rho_{n,1}^{(1)} = \chi_V \rho_{n,1,i}^{(1)}$$
$$= \chi_V \left\{ \rho_{n,1,f}^{(1)} \right. \qquad (7)$$
$$\left. + \frac{2}{3A_R} \tilde{\chi}_{\Delta\nu} \frac{\Delta\nu_l^{(1)}}{\nu_0} \left[ 1 + \tilde{B}_{\Delta\nu} \frac{\Delta\nu_l^{(1)}}{\nu_0} + \tilde{C}_{\Delta\nu} \left( \frac{\Delta\nu_l^{(1)}}{\nu_0} \right)^2 + \ldots \right] \right\}$$

This shows that to obtain an adequate estimate of $\rho_{n,ext}^{(0)}$ from a single assessment of $\Delta\nu_1^{(1)}$, not only access to the same entities as when the open system was analyzed, i.e. $A_R$, $\tilde{\chi}_{\Delta\nu}$, etc., but also the gas expansion factor, $\chi_V$, is needed.

*2. Double refractivity assessment – The use a second refractivity measurement for an automatic calibration of the system with respect to its ratio of volumes – A means for volume-independent assessments of the gas density*

Even if the gas expansion factor is not known beforehand, it is possible to provide a reliable assessment of $\rho_{n,ext}^{(0)}$ by performing a second measurement, preferably in rapid succession to the first. The justification and the methodology for this are as follows.

Following the single refractivity measurement, by evacuating and filling cavity 1 once more, the density of gas in this cavity, then denoted $\rho_{n,1}^{(2)}$, can be related to the density of gas in the external gas compartment after the 1st gas equilibration, $\rho_{n,ext}^{(1)}$, and, according to Eq. (6) thereby also the number density of molecules in cavity 1 after the first equilibration, $\rho_{n,1}^{(1)}$, through the gas expansion factor. i.e. by

$$\rho_{n,1}^{(2)} = \frac{1}{\chi_V} \rho_{n,ext}^{(1)} = \frac{1}{\chi_V} \rho_{n,1}^{(1)}, \qquad (8)$$

where the last step uses the fact that $\rho_{n,1}^{(1)} = \rho_{n,ext}^{(1)}$ from Eq. (6). This shows that $\chi_V$ can be assessed as

$$\chi_V = \frac{\rho_{n,1}^{(1)}}{\rho_{n,1}^{(2)}}, \tag{9}$$

where the two $\rho_{n,1}^{(i)}$ are to be assessed by the use of Eq. (1).

Plugging this into Eq. (6) yields

$$\rho_{n,ext}^{(0)} = \frac{[\rho_{n,1}^{(1)}]^2}{\rho_{n,1}^{(2)}}. \tag{10}$$

This implies that $\rho_{n,ext}^{(0)}$ can be directly assessed from the two measured shifts of the cavity mode addressed in cavity 1, $\Delta v_1^{(1)}$ and $\Delta v_1^{(2)}$, with no need to separately assess the gas expansion factor, and thereby with no requirement to assess the various compartment volumes.

The initial density of the external gas compartment can thereby be explicitly expressed in terms of the shifts of the beat frequencies by the use of Eq. (1) as

$$\rho_{n,ext}^{(0)} \approx \frac{\Delta v_1^{(1)}}{\Delta v_1^{(2)}} \rho_{n,1,f}^{(1)} + \frac{2}{3A_R} \tilde{\chi}_{\Delta v} \frac{\Delta v_1^{(1)}}{\Delta v_1^{(2)}} \frac{\Delta v_1^{(1)}}{v_0}$$
$$\times \left\{ 1 + \tilde{B}_{\Delta v} \left( \frac{2\Delta v_1^{(1)} - \Delta v_1^{(2)}}{v_0} \right) \right.$$
$$- \left( \tilde{B}_{\Delta v} \right)^2 \frac{\Delta v_1^{(2)}}{v_0} \left( \frac{\Delta v_1^{(1)} - \Delta v_1^{(2)}}{v_0} \right)$$
$$+ \tilde{C}_{\Delta v} \left[ 2 \left( \frac{\Delta v_1^{(1)}}{v_0} \right)^2 - \left( \frac{\Delta v_1^{(2)}}{v_0} \right)^2 \right]$$
$$+ \left. \frac{1}{\tilde{\chi}_{\Delta v}} \left( \frac{\rho_{n,1,f}^{(1)}}{\rho_{n,1,i}^{(1)}} - \frac{\rho_{n,1,f}^{(2)}}{\rho_{n,1,i}^{(2)}} \right) \right\}, \tag{11}$$

where we, in the first term for the residual gas density, which is considered to be a small entity, solely have retained the leading order of $\chi_V$, viz. $\Delta v_1^{(1)} / \Delta v_1^{(2)}$.

This shows that despite the fact that Eq. (7) contains the gas expansion factor, which is composed of two unknown volumes, $V_1$ and $V_{ext}$, the gas density in the external gas compartment, $\rho_{n,ext}^{(0)}$, can be assessed from finite-sized compartments by a double refractivity assessment with good accuracy, presumably by that by which $A_R$ and $\tilde{\chi}_{\Delta v}$ (or their ratio) can be assessed.

An analysis of this expression shows though that, in practice, not all terms in Eq. (11) need to be considered.[33] In particular, the $\tilde{B}_{\Delta v}^2$-term can in most cases be considered to be insignificant with respect to the $\tilde{B}_{\Delta v}$-term. Also the $\tilde{C}_{\Delta v}$-term can be considered to be negligible with respect to the $\tilde{B}_{\Delta v}$-term. In addition, in the $\tilde{\chi}_{\Delta v}$ entity, $\varepsilon^2$ can be neglected with respect to $1 - \varepsilon$ and $\eta$ can be neglected with respect to $\varepsilon$. In these cases, it is sufficient to assess $\rho_{n,ext}^{(0)}$ from the two shifts of the beat frequency measurements, $\Delta v_1^{(1)}$ and $\Delta v_1^{(2)}$, respectively, as

$$\rho_{n,ext}^{(0)} \approx \frac{\Delta v_1^{(1)}}{\Delta v_1^{(2)}} \rho_{n,1,f}^{(1)} + \frac{2}{3A_R} \frac{\Delta v_1^{(1)}}{\Delta v_1^{(2)}} \frac{\Delta v_1^{(1)}}{v_0} (1 - \varepsilon + 2\varsigma)$$
$$\times \left[ 1 + \tilde{B}_{\Delta v} \left( 2\frac{\Delta v_1^{(1)}}{v_0} - \frac{\Delta v_1^{(2)}}{v_0} \right) \right]. \tag{12}$$

This implies that if the system can be well characterized, i.e. by performing a measurement with a known gas density, by the use of either Eq. (11) or (12) and by keeping track of (e.g. measuring) the residual gas densities (as discussed above), the system can be used, together with either of these equations, to assess gas density with good accuracy under almost any condition.

Although the temperature dependence of the methodologies scrutinized in this work is estimated and discussed in some detail in our accompanying work,[25] it can here be concluded that the technique has a very small temperature dependence; since the polarizability (the $A_R$ entity) does not have any temperature dependence, and since the technique is assumed to be performed under drift free conditions, the only temperature dependence of these methodologies are provided through the instantaneous deformation of the spacer, $\varepsilon$, and the $\tilde{B}_{\Delta v}$-coefficient, which both provide small contributions to the assessed density.

The methodologies presented in the sections **B** and **C** above thus describe possible means to assess gas density at a given instance. It is obvious that any of the procedures described can be repeated after a given time. Since the precision of the technique is excellent, this will make it possible to detect tiny time dependent changes in the gas density. These methodologies are therefore not only useful for precise and accurate gas density measurements; they can also be utilized for applications where a change in gas density is the relevant entity, e.g. detection of small leaks of (possibly hazardous) gases in various environments. However, to be able to assess a leak rate given in terms of the number of molecules (or moles) per time unit, which is of special interest for characterization of standard reference gas leaks or when a passive (presumably unknown) leak is to be quantified, the aforementioned methodologies need to be slightly expanded.

**D. Methodology for assessment of leak rates – standard reference gas leaks**

As is shown in Fig. 5c, also when the leak rate of a standard reference gas leaks is to be assessed, which, for all practical reasons, can be considered to produce a constant leak rate, the generic FS-DFPC-OR system can preferably be directly connected to the closed external compartment to which a reference leak is connected. In this case, when a gas flow is to be assessed, a triple assessment methodology (comprising two sets of measurements performed in rapid succession, as described above, plus a third measurement at a certain time thereafter) should be performed.

*1. Time dependence of the gas density in the external gas compartment due to a standard reference gas leak*

The leak rate of a standard reference gas leak, $q_n^{SR}$, given in terms of number of molecules per second, can be assessed in terms of the rate of change of the number density of the gas in the external gas compartment as[34]

$$\frac{\partial \rho_{n,ext}(t)}{\partial t} = \frac{q_n^{SR}}{V_{ext}}. \tag{13}$$

This implies that the density in the external gas compartment will increase with time as

$$\rho_{n,ext}(t_j) = \rho_{n,ext}(t_i) + \frac{q_n^{SR}}{V_{ext}}(t_j - t_i). \tag{14}$$

where $t_i$ and $t_j$ ($> t_i$) are two arbitrarily chosen time instances.

*2. Means to assess the rate of a gas leak*

The rate of a gas leak can be determined by expanding the gas in the external gas compartment into cavity 1 a third time, a certain time after the other two assessments. However, since the density of gas in the external gas compartment will increase with time, to describe this, the time instances when the various assessments are performed need to be well defined. Let us assume that the three measurements takes place at times $t_1$, $t_2$, and $t_3$, respectively. This implies that the density in the measurement cavity directly after each of these time instances (i.e. after equalization of the conditions in the measurement cavity and the external gas compartment), $\rho_{n,1}^{(i)}(t_i)$, can be related to those in the external gas compartment, both after and before the gas equalization, by $\rho_{n,ext}^{(i)}(t_i)$ and $\rho_{n,ext}^{(i-1)}(t_i)$, by

$$\rho_{n,1}^{(1)}(t_1) = \rho_{n,ext}^{(1)}(t_1) = \frac{1}{\chi_V}\rho_{n,ext}^{(0)}(t_1),  \quad (15)$$

$$\rho_{n,1}^{(2)}(t_2) = \rho_{n,ext}^{(2)}(t_2) = \frac{1}{\chi_V}\rho_{n,ext}^{(1)}(t_2), \quad (16)$$

and

$$\rho_{n,1}^{(3)}(t_3) = \rho_{n,ext}^{(3)}(t_3) = \frac{1}{\chi_V}\rho_{n,ext}^{(2)}(t_3), \quad (17)$$

while the temporal development of the density in the external gas compartment in between these time instances can be written as

$$\rho_{n,ext}^{(1)}(t_2) = \rho_{n,ext}^{(1)}(t_1) + \frac{q_n^{SR}}{V_{ext}}(t_2 - t_1) \quad (18)$$

and

$$\rho_{n,ext}^{(2)}(t_3) = \rho_{n,ext}^{(2)}(t_2) + \frac{q_n^{SR}}{V_{ext}}(t_3 - t_2). \quad (19)$$

These five expressions constitute eight equations with which the six gas densities of the external compartment, i.e. those of the type $\rho_{n,ext}^{(j-1)}(t_j)$ and $\rho_{n,ext}^{(j)}(t_j)$, and the gas expansion factor can be eliminated. Doing so provides an expression for the leak rate, $q_n^{SR}$, in terms of the measurable entities, i.e. the various densities for cavity 1, i.e. $\rho_{n,1}^{(1)}(t_1)$, $\rho_{n,1}^{(2)}(t_2)$, and $\rho_{n,1}^{(3)}(t_3)$, and the differences between the various measurement instances, $t_2 - t_1$ and $t_3 - t_2$, that can be written as

$$q_n^{SR} = V_{ext} \frac{\rho_{n,1}^{(3)}(t_3)\rho_{n,1}^{(1)}(t_1) - \left[\rho_{n,1}^{(2)}(t_2)\right]^2}{\rho_{n,1}^{(2)}(t_2)(t_3 - t_2) - \rho_{n,1}^{(3)}(t_3)(t_2 - t_1)}, \quad (20)$$

where the various $\rho_{n,1}^{(i)}(t_i)$ are assessed by the use of Eq. (1) in terms of $\Delta v_1^{(i)}$.

For the case when the required accuracy of the leak rate is smaller than the relative contributions of the higher order terms, which typically are in the $10^{-4}$ range, and for the case when the two first assessments are made in rapid succession so that $(t_2 - t_1) \ll (t_3 - t_2)$, Eq. (20) can be simplified to

$$q_n^{SR} = \frac{2}{3A_R}\frac{V_{ext}}{t_3 - t_2}\frac{\Delta v_1^{(2)}}{v_0}\left(\frac{\Delta v_1^{(1)}\Delta v_1^{(3)}}{\Delta v_1^{(2)}\Delta v_1^{(2)}} - 1\right). \quad (21)$$

These expressions show that the leak rate from a standard reference gas leak can be determined with good accuracy by the use of three successive measurements. This time, not only knowledge about the same entities as when gas density is assessed is needed, but also $V_{ext}$ (or, since $V_{ext}$ is related to $V_1$, $\chi_V$ and $V_1$) needs to be known with the necessary accuracy.

### E. Methodology for assessment of changes of gas density – the leak rate through an orfice plate

When the leak rate of other types of gas leaks is to be assessed, a triple assessment methodology, similar to that used for when a standard reference leak is assessed, can be used.

*1. Time dependence of the gas density in the external gas compartment in the presence of an orfice plate*

For the case when a gas leak takes place through a small opening, it is suitable to model the flow as that thorough an orfice plate. It is customary to express the mass flow of gas through such a plate, $q_m^P$, (in units of kg/s) as [35]

$$q_m^P = CYA_O\sqrt{2\rho_{m,ext}(p_{ext} - p_{sur})}, \quad (22)$$

where $Y$ is an dimensionless expansibility factor of the gas (unity for an incompressible flow), $A_O$ the area of the orfice (m$^2$), $\rho_{m,ext}$ the mass density in the external gas compartment (in units of kg/m$^3$), and $p_{ext}$ and $p_{sur}$ the pressures in the external gas compartment and in the surrounding, respectively (both in Pa), where the former here is assumed to be larger than the latter. Furthermore, $C$ is a dimensionless constant that takes various properties of the gas and the orfice into account and that is specifically given or to be assessed for each leak. When this type of leak is to be characterized, it is often the product $CYA_O$ that is assessed.

Although presumably not considered for the first time, according to the authors' knowledge, there is no expression in the literature that describes the time dependence of the gas density in a closed compartment to (or from) which gas flows through such an orfice plate. An expression for this is therefore derived in the supplementary material. It is shown there that the density of molecules in the external compartment, $\rho_{n,ext}$, will, until the gas densities in the two compartments have been equilibrated, decrease with time as

$$\rho_{n,ext}^{(i)}(t_j) = \rho_{n,ext}^{(i)}(t_i)\Phi_i e^{-(t_j - t_i)/t_0}\left[1 + \frac{\rho_{n,sur}}{\rho_{n,ext}^{(i)}(t_i)}\frac{\Phi_i}{4}e^{(t_j - t_i)/t_0}\right]^2, \quad (23)$$

where $\rho_{n,sur}$ is the gas density in the surrounding, to which the gas leak leaks, $\Phi_i$ is a constant that depends on the initial conditions of the system, given by

$$\Phi_i = \frac{1}{2}\left[1 + \sqrt{1 - \frac{\rho_{n,sur}}{\rho_{n,ext}^{(i)}(t_i)}} - \frac{\rho_{n,sur}}{2\rho_{n,ext}^{(i)}(t_i)}\right]$$
$$\approx 1 - \frac{1}{2}\frac{\rho_{n,sur}}{\rho_{n,ext}^{(i)}(t_i)} - \frac{1}{16}\left(\frac{\rho_{n,sur}}{\rho_{n,ext}^{(i)}(t_i)}\right)^2 + \ldots \quad (24)$$

where the latter series expansion is valid when $\rho_{n,sur} \ll \rho_{n,ext}^{(i)}(0)$, and $t_0$ is a characteristic leakage time of the system that is defined as[36]

$$t_0 = \frac{V_{ext}}{CYA_O}\sqrt{\frac{M_x}{2N_AkT}}. \quad (25)$$

*2. Means to assess the leak rate from an orfice plate*

As for the case with the standard reference gas leak, by expanding the gas in the external gas compartment into cavity 1 three times, at times $t_1$, $t_2$, and $t_3$, respectively, the density in the measurement cavity directly after each of these time instances (i.e. after equalization of the conditions in the measurement cavity and the external gas compartment) can be related to those in the external gas compartment by the Eqs. (15) - (17). The expressions for the time development of the gas in the external gas compartment will then read

$$\rho_{n,ext}^{(1)}(t_2) = \rho_{n,ext}^{(1)}(t_1)\Phi_1 e^{-(t_2 - t_1)/t_0}\left[1 + \frac{\rho_{n,sur}}{\rho_{n,ext}^{(1)}(t_1)}\frac{\Phi_1}{4}e^{(t_2 - t_1)/t_0}\right]^2 \quad (26)$$

and

$$\rho_{n,ext}^{(2)}(t_3) = \rho_{n,ext}^{(2)}(t_2)\Phi_2 e^{-(t_3 - t_2)/t_0}\left[1 + \frac{\rho_{n,sur}}{\rho_{n,ext}^{(2)}(t_2)}\frac{\Phi_2}{4}e^{(t_3 - t_2)/t_0}\right]^2. \quad (27)$$

Again, as for the case with the standard reference gas leak, the Eqs. (15) - (17), together with the expressions for the time development of the gas in the external gas compartment, in this case the Eqs. (26) and (27), can be used [by eliminating the gas densities of the external compartment, i.e. those of the type $\rho_{n,ext}^{(j-1)}(t_j)$ and $\rho_{n,ext}^{(j)}(t_j)$ ] to assess the leak rate of the system.

For the case when the leak is so small so that the change in density in the external gas compartment between the two first assessments can be neglected, for which the gas expansion factor is given by Eq. (9), it is possible to simplify the set of expressions [Eqs. (15) - (17) together with the Eqs. (26) and (27)] to

$$\rho_{n,1}^{(3)}(t_3) = \frac{\rho_{n,ext}^{(2)}(t_3)}{\chi_V}, \quad (28)$$

where the $\rho_{n,ext}^{(2)}(t_2)$ entity in the expression for $\rho_{n,ext}^{(2)}(t_3)$, given by Eq. (27), is, according to Eq. (16), given by the measurable $\rho_{n,1}^{(2)}(t_2)$ entity. In this case, it is possible to solve Eq. (28), with the use of Eq. (27), for $t_0$. Doing so yields

$$t_0 = \frac{t}{\ln\left\{4\Phi_2 \frac{\rho_{n,ext}^{(i)}(t_i)}{\rho_{n,sur}}\left[2\overline{CE}\left(1+\sqrt{1-\frac{1}{\overline{CE}}}\right)-1\right]\right\}}, \quad (29)$$

where $\overline{CE}$ is given by Eq. (S7) in the supplementary material, which, after series expansion of its denominator, can be expressed as

$$\overline{CE} = \frac{2}{3A_R\rho_{n,sur}}\frac{\Delta\nu_1^{(2)}\Delta\nu_1^{(3)}}{\Delta\nu_1^{(1)}\nu_0}$$
$$\times\left\{1+\tilde{B}_{\Delta\nu}\left(\frac{\Delta\nu_1^{(1)}-\Delta\nu_1^{(2)}+\Delta\nu_1^{(3)}}{\nu_0}\right)\right.$$
$$-\left(\tilde{B}_{\Delta\nu}\right)^2\left(\frac{\Delta\nu_1^{(1)}-\Delta\nu_1^{(2)}}{\nu_0}\right)\left(\frac{\Delta\nu_1^{(2)}-\Delta\nu_1^{(3)}}{\nu_0}\right) \quad (30)$$
$$+\tilde{C}_{\Delta\nu}\left[\left(\frac{\Delta\nu_1^{(1)}}{\nu_0}\right)^2-\left(\frac{\Delta\nu_1^{(2)}}{\nu_0}\right)^2+\left(\frac{\Delta\nu_1^{(3)}}{\nu_0}\right)^2\right]$$
$$\left.+\frac{1}{\tilde{\chi}_{\Delta\nu}}\left[\frac{\rho_{n,1,f}^{(1)}}{\rho_{n,1,i}^{(1)}}-\frac{\rho_{n,1,f}^{(2)}}{\rho_{n,1,i}^{(2)}}+\frac{\rho_{n,1,f}^{(3)}}{\rho_{n,1,i}^{(3)}}\right]\right\}.$$

These expressions thus show that, from a triple refractivity assessment, it is possible to assess the characteristic leakage time of the system, $t_0$. As for the case when $\rho_{n,ext}^{(0)}$ is assessed by the use of Eq. (11), it is possible to conclude that not all terms in the curly brackets in Eq. (30) need to be considered in practice. In particular, the $\tilde{B}_{\Delta\nu}^2$-term can in most cases be considered to be insignificant with respect to the $\tilde{B}_{\Delta\nu}$-term. Also the $\tilde{C}_{\Delta\nu}$-term can be considered to be negligible with respect to the $\tilde{B}_{\Delta\nu}$-term. The remaining terms can also often be neglected, however, depending on the actual conditions. It can then be concluded, e.g. from Eq. (S2), that the actual leakage rate from an orfice plate, in terms of number of molecules per second, which, in fact is a function of time, then denoted $q_n^{PGL}(t)$, can be assessed from this as

$$q_n^{PGL}(t) = \frac{V_{ext}}{t_0}\sqrt{\rho_{n,ext}^{(2)}(t)\left[\rho_{n,ext}^{(2)}(t)-\rho_{n,sur}\right]}, \quad (31)$$

where $\rho_{n,ext}^{(2)}(t)$, if high accuracy is needed, is given by Eq. (27), while, for the cases when $t/t_0 \ll 1$, it is given by the last expression in the second footnote in the supplementary material.

Note that this assessment can be done with very little influence of drifts of temperature. The largest temperature dependence might be the leak itself.

## VI. SUMMARY AND CONCLUSIONS

It has previously been concluded that although Fabry-Perot Cavity based Optical Refractometry (FPC-OR) has a high potential for assessment of gas density, in particular an exceptional precision, its performance is often, in practice, limited by drifts in the FP-cavity. Specifically, its accuracy can be restricted by concealed non-linear responses. The analysis performed in our accompanying works[24, 25] show though that as long as measurements can be performed under so called "drift-free" conditions, the FPC–OR technique has an extraordinary potential for assessment of gas density.

An experimental characterization of a typical Dual FPC-OR system (DFPC-OR), as reported in section 4, indicates, by an Allan-Werle plot, that measurements are affected by drifts if they are performed for longer periods of time. This implies that drift free assessments can be effectuated if the measurements are performed under sufficiently short time periods. Therefore, to alleviate the aforementioned types of problems, and to bring out the most of the DFPC-OR technique, we advocate in this work that DFPC-OR can be performed with a minimum of influence of drifts if it could be performed by *fast switching* DFPC-OR (FS-DFPC-OR), which implies that the assessments of the decrease of the refractivity of the gas during evacuation of the measurement cavity should be made under such short time periods that any possible drifts of the length of the cavity can be disregarded.[37]

Suggested methodologies for assessment of gas densities under drift free conditions, in both open systems and closed compartments, and gas flows, from both standard reference gas leaks and orfice plates, by FS-DFPC-OR have been worked out in some detail. Since the methodologies are based upon measurements under drift free conditions, the major causes for degradation of its precision and accuracy have been eliminated.

In particular, it is outlined how gas density can be assessed from closed compartments without need for any specific assessment of individual gas volumes or gas expansion factors. A methodology that consists of two evacuation measurements in rapid succession is presented. While the first one serves the purpose of assessing the density of the gas that has been transferred into the measurement cavity by the gas equilibration process, the second is used to automatically calibrate the system with respect to the relative volumes of the measurement cavity and the external compartment. By this, it is suggested that gas densities in closed, finite-sized compartments can be assessed by FPC-OR by means of volume-independent assessments.

The methodology for assessments of flow rates comprise of a triple cavity evacuation assessment, consisting of two sets of measurements performed in rapid succession, as for the gas density assessment, supplemented by a third measurement at a certain time thereafter.

It has previously been stated that the technique has a very small temperature dependence. Although the precision, accuracy, and temperature dependence of the methodologies scrutinized in this work are estimated and discussed in some detail in an accompanying work,[25] it has been explicated in this work why this is the case.[38] This implicates that the system can be used to assess gas density with the same accuracy under a variety of conditions and that it is particularly useful under non-temperature stabilized ones.

As is further alluded to in our accompanying work,[25] when fully characterized, the use of the fast-switching methodologies is expected to improve of the performance of FPC-OR so that, whence properly characterized, it can potentially be limited by the locking of the laser to cavity or the reproducibility in the gas handling process.

In summary, together with our accompanying works, which derive general and explicit expressions for how a measured change in laser frequency is related to the density of a gas[24] and provide an analysis of the estimated precision, accuracy, and temperature dependence of DF-DFPC-OR, which shows that the technique has an extraordinary potential for assessments of gas density,[25] we consider that we, by this work, have laid the foundation for novel methodologies for DF-FPC-OR in general, and DF-DFPC-OR in particular, that are based on assessments of rapid changes of gas densities in the measurement cavity, FS-DFPC-OR„ that can be used for precise and accurate assessment of gas number density and gas flows under a variety of conditions, in particular non-temperature stabilized ones. It is hoped that this work can serve as a basis for future constructions of instrumentations that can benefit the most from the extraordinary power of DF-DFPC-OR for gas density assessments.

**SUPPLEMENTARY MATERIAL**

See the supplementary material for information about the time dependence of the gas density in the external gas compartment due to an orfice plate.

**Acknowledgment**

This research was supported by the EMPIR initiative, which is co-founded by the European Union's Horizon 2020 research and innovation program and the EMPIR Participating States; the Swedish Research Council (VR), project numbers 621-2011-4216 and 621-2015-04374; Umeå University program VFS (Verifiering för samverkan), 2016:01; and Vinnova Metrology Programme, project number 2015-0647 and 2014-06095.

**References**

1. P. J. Mohr, D. B. Newell and B. N. Taylor, Rev. Mod. Phys. **88** (3) (2016).
2. L. R. Pendrill, Metrologia **41** (2), S40-S51 (2004).
3. A. Picard and H. Fang, Metrologia **39** (1), 31-40 (2002).
4. L. R. Pendrill, Metrologia **25** (2), 87-93 (1988).
5. H. Fang and P. Juncar, Rev. Sci. Instrum. **70** (7), 3160-3166 (1999).
6. H. Fang and P. Juncar, in *Interferometry '99: Applications*, edited by W. P. O. Juptner and K. Patorski (1999), Vol. 3745, pp. 189-195.
7. H. Fang, A. Picard and P. Juncar, Rev. Sci. Instrum. **73** (4), 1934-1938 (2002).
8. A. Picard and H. Fang, IEEE T. Instrum. Meas. **52** (2), 504-507 (2003).
9. E. Hedlund and L. R. Pendrill, Meas. Sci. Technol. **17** (10), 2767-2772 (2006).
10. E. Hedlund and L. R. Pendrill, Meas. Sci. Technol. **18** (11), 3661-3663 (2007).
11. P. Egan and J. A. Stone, Appl. Opt. **50** (19), 3076-3086 (2011).
12. I. Silander, M. Zelan, O. Axner, F. Arrhen, L. Pendrill and A. Foltynowicz, Meas. Sci. Technol. **24** (10), 105207 (2013).
13. P. F. Egan, J. A. Stone, J. H. Hendricks, J. E. Ricker, G. E. Scace and G. F. Strouse, Opt. Lett. **40** (17), 3945-3948 (2015).
14. P. F. Egan, J. A. Stone, J. E. Ricker and J. H. Hendricks, Rev. Sci. Instrum. **87** (5), 053113 (2016).
15. H. J. Achtermann, G. Magnus and T. K. Bose, J. Chem. Phys. **94** (8), 5669-5684 (1991).
16. J. W. Schmidt, R. M. Gavioso, E. F. May and M. R. Moldover, Phys. Rev. Lett. **98** (25), 254504 (2007).
17. J. A. Stone and A. Stejskal, Metrologia **41** (3), 189-197 (2004).
18. M. Andersson, L. Eliasson and L. R. Pendrill, Appl. Opt. **26** (22), 4835-4840 (1987).
19. N. Khelifa, H. Fang, M. Himbert and P. Juncar, presented at the Ninth International School on Quantum Electronics: Lasers-Physics and Applications, Varna, Bulgaria, 1996.
20. N. Khelifa, H. Fang, J. Xu, P. Juncar and M. Himbert, Appl. Opt. **37** (1), 156-161 (1998).
21. H. Fang, M. Himbert, N. Khelifa, P. Juncar and L. R. Pendrill, presented at the 2000 Conference on Precision Electromagnetic Measurements Digest, Sydney, Australia, 2000.
22. L. P. Yan, B. Y. Chen, E. Z. Zhang, S. H. Zhang and Y. Yang, Rev. Sci. Instrum. **86** (8), 085111 (2015).
23. J. H. Hendricks, G. F. Strouse, J. E. Ricker, D. A. Olson, G. E. Scace, J. A. Stone and P. F. Egan, US Patent No. US20160018280 A (Mar. 20 2015).
24. O. Axner, I. Silander, T. Hausmaninger and M. Zelan, arXiv:1704.01187.
25. M. Zelan, I. Silander, T. Hausmaninger and O. Axner, arXiv:1704.01185.
26. D. W. Allan, Proceedings of the IEEE **54** (2), 221-230 (1966).

27. Note that in comparison with ultra-stable cavities used for frequency metrology, see e.g. [M. Notcutt, L. S. Ma, J. Ye and J. L. Hall, Opt. Lett. 30 (14), 1815-1817 (2005)], a relative precision of 2 × 10$^{13}$ is not particular impressive. This performance of the system characterized here does therefore not reflect the maximum stability that can be achieved using this technique.

28. It can be noticed that the analysis in our accompanying work, Eq. (34) in Ref. [25], shows that if a minimum fluctuation of the beat frequency of 36 Hz would correspond to a temperature drift, it would, for a spacer material with an $\alpha$ of 10$^{-7}$ K$^{-1}$, represent a drift during the evacuation of 200 μK. Since this value was achieved for a measurement time of 0.25 s, this would correspond to a temperature drift rate of ~1 mK/s. However, the system is assumed to have a significantly smaller temperature drift than this (rather 2 mK/hour, corresponding to 0.5 μK/s). This indicates that this minimum in the fluctuations of the beat frequency is presumably not due to a temperature drift of the cavity spacer material. Although not further scrutinized in detail here, possible causes are relaxations in the material, which previously have been noticed [13], or, for this particular case, since the mirrors were glued onto the cavity spacer material, thermal expansion of the glue. Irrespective of the cause of the minimum fluctuations of the beat frequency, since measurements for the FS-DFPC-OR methodology are solely performed under a short time, this shows that the requirement on the temperature stability during the evacuation of the cell is not particular severe. An obvious means of improvement of the performance of an DFPC-OR would then be to fasten the mirrors by optical contacting, see Refs. [11, 13, 14, 20] and [G. Kalkowski, S. Risse, C. Rothhardt, M. Rohde and R. Eberhardt, in Optical Manufacturing and Testing Ix, edited by J. H. Burge, O. W. Fahnle and R. Williamson (2011), Vol. 8126].

29. We have here used the notation that the superscript (i) denotes the number of times the gas has been equilibrated from an external gas compartment into the measurement cavity, which, in turn, is assumed to be evacuated prior to each gas transfer.

30. Note that even if such an expansion of the gas will decrease the temperature, the final density in the cavity is not affected by this temperature decrease.

31. This also implies that the density in the external cavity has been decreased by $[V_1/(V_{ext}+V_1)]\rho_{ext}^{(0)}$, which alternatively is expressible as $(1-X_V^{-1})\rho_{ext}^{(0)}$.

32. A similar expression for the case when relocking is used can be derived. However, due to space constraints, such expressions will not be explicitly given in the rest of the paper. If needed though, the reader should be able to derive the expressions that for this case correspond to those given in this work for the no relocking situation.

33. For the case with $V_1=V_{ext}/10$, $X_V$ becomes 1.1. This implies that $\Delta \nu_1^{(2)}$ is 91% of $\Delta \nu_1^{(1)}$. Since, at STP, $\Delta \nu_1^{(1)}/\nu_0$ is around 3 × 10$^{-4}$, $[\Delta \nu_1^{(1)}-\Delta \nu_1^{(2)}]/\nu_0$ becomes 2.7 × 10$^{-5}$. For N$_2$, $B_{\Delta \nu}$ is around 0.81 for the 632 nm wavelength. This implies that the first non-linear term in Eq. (11), the one proportional to $B_{\Delta \nu}$, is 2.7 × 10$^{-4}$. The next terms is 4.8 × 10$^{-10}$, and can therefore be neglected. The $C_{\Delta \nu}$ coefficient has a value of -1.28. This term has thereby a value of -1.35 × 10$^{-7}$. Under the condition that the evacuation can be made so that $\rho_{n,1,f}^{(i)}$ is a small fraction of $\rho_{n,1,f}^{(i)}$, e.g. less than 10$^{-5}$, and that the evacuation process can be made with a repeatability of at least a few ‰ (so $\rho_{n,1,i}^{(i)}$ and $\rho_{n,1,f}^{(i)}$ differ by this amount), the last term can be assumed to be < $10^{-7}$. This implies that in many cases, also the two last terms can be neglected.

[34] For the case when the gas leak is given as a mass flow rate, i.e. in terms of kg/s, denoted $q_m$, it can be related to the flow rate expressed in terms of number of molecules per second, $q_n$, according to $q_n=(N_A/M_x)q_m$ where $N_A$ is Avogadro's number (the number of molecules per mole) and $M_x$ the molar mass (in kg/mol).

[35] R. H. Perry and D. W. Green, (McGraw-Hill, 2007).

[36] For $N_2^{STP}$: $M_x$ = 1.25 kg/m$^3$, which implies that $[M_x/(2N_AkT)]^{1/2}$ = 0.01659 s/m. If $V_{ext}$ is 1 dm$^3$, i.e. $10^{-3}$ m$^3$, and $A_0$ is circular with a radius of 1 μm, i.e. having a value of 3.14 × $10^{-12}$ m$^2$, $Y$ is 1, and $C$ = 1, then $t_0$ becomes 5.3 × $10^6$ s. For the case with an aperture with a diameter of 100 μm, $t_0$ becomes 530 s.

[37] Although the cavity spacer used in this work had not been exposed to any thermalization or curing to decrease internal relaxations, whereby its drifts set in already after averaging times above 0.5 s, we do not see this time as universal limit for drift-free DFPC-OS measurements. On the contrary, it is presumed that systems in which drifts set in at significantly longer averaging times can be constructed; one such has been demonstrated by Khelifa et al. which has demonstrated white-noise limited behavior to at least 100 s.[20] This implies that the requirement for a measurement to be considered to be "fast switching" can, in practice, be that it is made during time periods that are shorter than several seconds or several tens of seconds.

[38] The main reasons are that refractivity has solely a weak non-linear dependence on gas density (provided though a virial $B$-coefficient) and that the leading linear dependence of reactivity on density, which is mediated though the polarizability (the $A_R$ entity), does not have any temperature dependence. Hence, the main contributions to the temperature dependence of the technique are primarily given by the virial $B$-coefficient and the instantaneous deformation of the spacer, $\varepsilon$.

# Supplementary material to "Fast switching dual Fabry-Perot cavity optical refractometry – methodologies for accurate assessment of gas density"

Isak Silander,[1] Thomas Hausmaninger,[1] Michael Bradley,[1] Martin Zelan[2,a)], and Ove Axner[1,a)]

[1]*Department of Physics, Umeå University, SE-901 87 Umeå, Sweden*
[2]*Measurement Science and Technology, RISE Research Institutes of Sweden, SE-501 15 Borås, Sweden*

## Part 1 – Time dependence of the gas density in the external gas compartment due to an orfice plate

Equation (22) provides the flow of gas though an orfice plate in terms of the mass flow of gas through an orfice plate (in units of kg/s) and the pressure of the gases (in Pa). By converting the mass density, $\rho_{m,ext}$, and the pressures, $P_{ext}$ and $P_{sur}$, to number densities, denoted $\rho_{n,ext}$ and $\rho_{n,sur}$, where for the former one $\rho_{n,ext} = (N_A / M_x)\rho_{m,ext}$ is used, while for the latter ones the ideal gas law, i.e. $p_{ext} = kT\rho_{n,ext}$ and $p_{sur} = kT\rho_{n,sur}$ is used, respectively, and where $\rho_{n,sur}$ is the number density of molecules in the surrounding (to which there is a leak), it is possible, for the cases when $\rho_{n,ext} > \rho_{n,sur}$, to express the number flow density, $q_n^P$, as

$$q_n^P = C\sqrt{\frac{2N_A kT}{M_x}} Y A_O \sqrt{\rho_{n,ext}(\rho_{n,ext} - \rho_{n,sur})}. \quad (S1)$$

The density of molecules in the external compartment, $\rho_{n,ext}^{(i)}$, will then decrease with time in a manner given by the differential equation

$$\frac{\partial \rho_{n,ext}^{(i)}(t)}{\partial t} = -\frac{q_n^P}{V_{ext}} = -\frac{1}{t_0}\sqrt{\rho_{n,ext}^{(i)}(t)\left[\rho_{n,ext}^{(i)}(t) - \rho_{n,sur}\right]}, \quad (S2)$$

where $t_0$ is a characteristic leakage time of the system that is given by

$$t_0 = \frac{V_{ext}}{CYA_O}\sqrt{\frac{M_x}{2N_A kT}}. \quad (S3)$$

It can be shown that, for times up to $t'$, which is the time it takes for the system to reach $\rho_{n,ext}^{(i)}(t') = \rho_{n,sur}$,[S1] Eq. (S2) has the solution

$$\rho_{n,ext}^{(i)}(t_j) = \rho_{n,ext}^{(i)}(t_i)\Phi_i e^{-(t_j-t_i)/t_0}\left[1 + \frac{\rho_{n,sur}}{\rho_{n,ext}^{(i)}(t_i)}\frac{\Phi_i}{4}e^{(t_j-t_i)/t_0}\right]^2, \quad (S4)$$

where

$$\Phi_i = \frac{1}{2}\left[1 + \sqrt{1 - \frac{\rho_{n,sur}}{\rho_{n,ext}^{(i)}(t_i)} - \frac{\rho_{n,sur}}{2\rho_{n,ext}^{(i)}(t_i)}}\right]$$
$$\approx 1 - \frac{1}{2}\frac{\rho_{n,sur}}{\rho_{n,ext}^{(i)}(t_i)} - \frac{1}{16}\left(\frac{\rho_{n,sur}}{\rho_{n,ext}^{(i)}(t_i)}\right)^2 + ... \quad (S5)$$

where $t_i$ and $t_j$ ($> t_i$) are two arbitrarily chosen time instances and where the latter series expansion is valid when $\rho_{n,sur} \ll \rho_{n,ext}^{(i)}(t_i)$.[S2]

---

[S1] The time $t'$ is given by

$$t' = t_0 \ln\left[\frac{4\rho_{n,ext}(0)\Phi_i}{\rho_{n,sur}}\right] \approx t_0 \ln\left[\frac{4\rho_{n,ext}(0)}{\rho_{n,sur}}\right]$$

where the latter again is valid when $\rho_{n,sur} \ll \rho_{n,ext}(0)$. This shows that when the gas density in the surrounding is significantly smaller than in the external gas compartment, $t'$ becomes a few times $t_0$, e.g. when $\rho_{n,1}(0) = 100\rho_{n,sur}$, $t'$ becomes $6t_0$, while for $\rho_{n,1}(0) = 10\rho_{n,sur}$, $t'$ becomes $3.7t_0$.

Solving Eq. (28) for $t_0$ yields, after some work,

$$t_0 = \frac{t}{\ln\left\{4\Phi_2\frac{\rho_{n,ext}^{(i)}(t_i)}{\rho_{n,sur}}\left[2\overline{CE}\left(1 + \sqrt{1 - \frac{1}{\overline{CE}}}\right) - 1\right]\right\}}. \quad (S6)$$

where the latter series expansion is valid when $\rho_{n,sur} \ll \rho_{n,ext}^{(2)}(0)$, and where

$$\overline{CE} =$$
$$= \frac{2}{3A_R\rho_{n,sur}}\frac{\Delta v_1^{(3)}}{v_0}\left[1 + \tilde{B}_{\Delta v}\frac{\Delta v_1^{(3)}}{v_0} + \tilde{C}_{\Delta v}\left(\frac{\Delta v_1^{(3)}}{v_0}\right)^2 + \frac{1}{\tilde{\chi}_{\Delta v}}\frac{\rho_{n,1,f}^{(3)}}{\rho_{n,1,i}^{(3)}}\right]\chi$$

$$= \frac{2}{3A_R\rho_{n,sur}}\frac{\Delta v_1^{(1)}\Delta v_1^{(3)}}{\Delta v_1^{(2)}v_0}$$
$$\times\frac{\left[1 + \tilde{B}_{\Delta v}\frac{\Delta v_1^{(1)}}{v_0} + \tilde{C}_{\Delta v}\left(\frac{\Delta v_1^{(1)}}{v_0}\right)^2 + \frac{1}{\tilde{\chi}_{\Delta v}}\frac{\rho_{n,1,f}^{(1)}}{\rho_{n,1,i}^{(1)}}\right]}{\left[1 + \tilde{B}_{\Delta v}\frac{\Delta v_1^{(2)}}{v_0} + \tilde{C}_{\Delta v}\left(\frac{\Delta v_1^{(2)}}{v_0}\right)^2 + \frac{1}{\tilde{\chi}_{\Delta v}}\frac{\rho_{n,1,f}^{(2)}}{\rho_{n,1,i}^{(2)}}\right]}$$
$$\times\left[1 + \tilde{B}_{\Delta v}\frac{\Delta v_1^{(3)}}{v_0} + \tilde{C}_{\Delta v}\left(\frac{\Delta v_1^{(3)}}{v_0}\right)^2 + \frac{1}{\tilde{\chi}_{\Delta v}}\frac{\rho_{n,1,f}^{(3)}}{\rho_{n,1,i}^{(3)}}\right]$$

$$= \frac{2}{3A_R\rho_{n,sur}}\frac{\Delta v_1^{(2)}\Delta v_1^{(3)}}{\Delta v_1^{(1)}v_0}$$
$$\times\left\{1 + \tilde{B}_{\Delta v}\left(\frac{\Delta v_1^{(1)} - \Delta v_1^{(2)} + \Delta v_1^{(3)}}{v_0}\right)\right.$$
$$- (\tilde{B}_{\Delta v})^2\left(\frac{\Delta v_1^{(1)} - \Delta v_1^{(2)}}{v_0}\right)\left(\frac{\Delta v_1^{(2)} - \Delta v_1^{(3)}}{v_0}\right)$$
$$+ \tilde{C}_{\Delta v}\left[\left(\frac{\Delta v_1^{(1)}}{v_0}\right)^2 - \left(\frac{\Delta v_1^{(2)}}{v_0}\right)^2 + \left(\frac{\Delta v_1^{(3)}}{v_0}\right)^2\right]$$
$$\left.+ \frac{1}{\tilde{\chi}_{\Delta v}}\left[\frac{\rho_{n,1,f}^{(1)}}{\rho_{n,1,i}^{(1)}} - \frac{\rho_{n,1,f}^{(2)}}{\rho_{n,1,i}^{(2)}} + \frac{\rho_{n,1,f}^{(3)}}{\rho_{n,1,i}^{(3)}}\right]\right\},$$

(S7)

---

[S2] An alternative view of the decay of the gas in the external gas compartment due to the leak is to series expand the full expression for the time development of the gas density in the external gas compartment, Eq. (S4). This can be suitable whenever $t/t_0 \ll 1$, i.e. for small leaks or short waiting times. In this case, $\rho_{n,ext}^{(i)}(t)$ can be written as

$$\rho_{n,ext}^{(i)}(t_j) = \rho_{n,ext}^{(i)}(t_i)$$
$$\times\left[1 - \sqrt{1 - \frac{\rho_{n,sur}}{\rho_{n,ext}^{(i)}(t_i)}}\frac{(t_j-t_i)}{t_0} + \frac{1}{2}\left(1 - \frac{1}{2}\frac{\rho_{n,sur}}{\rho_{n,ext}^{(i)}(t_i)}\right)\left(\frac{t_j-t_i}{t_0}\right)^2 + ...\right]$$

The first time dependent term in this expression thus represents the leak rate when it is treated as constant rate.

where in the first step $\chi_V$ has been taken as given by Eq. (9). In the last step a series expansion of the denominator is performed and terms that are in the order of $(\Delta \nu_1^{(i)} / \nu_0)^3$ or higher have been neglected.